# Leveraging Legacy Data to Accelerate Materials Design via Preference Learning

Xiaolin Sun[1], Zhufeng Hou[2,3], Masato Sumita[4,5], Shinsuke Ishihara[4], Ryo Tamura[1,2,4,5], Koji Tsuda[1,2,5]

[1]Graduate School of Frontier Sciences, The University of Tokyo, 5-1-5 Kashiwa-no-ha, Kashiwa, Chiba 277-8561, Japan
[2]Research and Services Division of Materials Data and Integrated System, National Institute for Materials Science, 1-2-1 Sengen, Tsukuba, Ibaraki 305-0047, Japan
[3]State Key Laboratory of Structural Chemistry, Fujian Institute of Research on the Structure of Matter, Chinese Academy of Sciences (CAS), Fuzhou, 350002 Fujian, P.R. China
[4]International Center for Materials Nanoarchitectonics (WPI-MANA), National Institute for Materials Science, 1-1 Namiki, Tsukuba, Ibaraki 305-0044, Japan
[5]Center for Advanced Intelligence Project, RIKEN, 1-4-1 Nihombashi, Chuo-ku, Tokyo 103-0027, Japan

E-mail: tsuda@k.u-tokyo.ac.jp

## Abstract

Machine learning applications in materials science are often hampered by shortage of experimental data. Integration with legacy data from past experiments is a viable way to solve the problem, but complex calibration is often necessary to use the data obtained under different conditions. In this paper, we present a novel calibration-free strategy to enhance the performance of Bayesian optimization with preference learning. The entire learning process is solely based on pairwise comparison of quantities (i.e., higher or lower) in the same dataset, and experimental design can be done without comparing quantities in different datasets. We demonstrate that Bayesian optimization is significantly enhanced via addition of legacy data for organic molecules and inorganic solid-state materials.

## 1. Introduction

A substantial amount of materials data are accumulated in public databases [1–3], and machine-learning-based design of materials is increasingly common in recent years [4,5]. The problem of materials design is mathematically formulated as a black-box optimization problem, where a large number of candidates are available and the goal is to find the candidate with best target property via a minimum number of observations. In Bayesian optimization [6], one of the most prominent methods of black-box optimization, a next candidate to observe is chosen using a Bayesian surrogate model trained with observed candidates. Gaussian process [7] is one of most frequently used surrogate models that provides prediction together with uncertainty quantification. The next candidate is chosen such that the chance of getting beyond the current best candidate is maximized.

Despite progress in materials informatics, machine learning often yields poor results due to shortage of experimental data [8]. The problem may be solved by augmenting the current dataset with a *legacy dataset* in public databases or private repositories. However, even if a dataset about a similar experiment is found, direct mixing often leads to poor results, because the experiment in the past was done with different instruments and conditions. Calibration of quantities is hard due to shortage of data. To make things worse, the conditions in past experiments are often poorly documented or completely unknown. Also, in materials design, the difficulty of making most of legacy data depends on the overlap between candidate materials and examples in legacy data (Figure 1). If all the candidate materials are included in legacy examples, legacy data would provide plenty of information about experimental design (Figure 1, right). With small overlap, it may be difficult to accelerate the search (Figure 1, left).

In this paper, we propose a new calibration-free strategy of data integration without comparing the quantities in different datasets. Figure 2 illustrates our basic idea. First of all, each dataset is described as a set of pairwise relationships. Pairwise comparison is done for every pair of target values and the outcome is summarized as a set of 'larger-than' relationships. Then, a Bayesian surrogate model is learned from the two sets of pairwise relationships only. As a result, the learned model has a value range completely different from the original datasets, but it can still be used to select candidates with Bayesian optimization. One can use any preference learning method, but we employed Gaussian process-based method by Chu and Ghahramani [9] in this paper.

In benchmarking our method, we consider two types of materials search problems. First, we search for organic molecules with longer absorption wavelength [4]. Bayesian optimization is applied to 14 candidate compounds whose absorption wavelength is known experimentally. As the legacy data, we employed TD-DFT to compute the wavelengths of 90 compounds including the 14 candidates. We performed several computational experiments with different degrees of overlap and found significant search acceleration in all cases including no-overlap. Second, an oxide with the largest bandgap is sought from 194 candidates [3]. Similar successful results were obtained with a legacy dataset of 2142 examples. Overall, preference learning was effective in exploiting information in legacy data and may serve as a new tool of data integration in a wide range of materials science problems.

## 2. Methods

A set of candidate materials is represented as $\{z_i\}_{i=1,\ldots,N}$, where $z_i \in \mathcal{R}^d$ is a vector of descriptors. The corresponding values of target property are represented as $\{y_i\}_{i=1,\ldots,N}$. They are initially unknown and revealed by observation. Let us assume that $k$ observations are already made $Z = \{(z_i, y_i)\}_{i=1,\ldots,k}$ and we would like to choose a next candidate. In addition, we have a legacy dataset $Z' = \{(z'_i, y'_i)\}_{i=1,\ldots,k'}$ at hand. Before merging the datasets, each one is converted to *preferences*. If $y_i > y_j$, we denote $z_i \succ z_j$, i.e., $z_i$ is preferred over $z_j$. After comparing all pairs, $Z$ and $Z'$ are converted to preference sets of size $\frac{k(k-1)}{2}$ and $\frac{k'(k'-1)}{2}$, respectively. A Gaussian process is trained from the merged preference set, and subsequently used to rank the remaining candidates for next observation. Note that no comparison is made across the two datasets.

## 2.1 Gaussian process preference learning

In this section, we briefly review the preference learning method by Chu and Ghahramani [9]. For notational simplicity, all descriptor vectors in $Z \cup Z'$ are redefined as $X = \{x_i\}_{i=1,\ldots,n}$. Let $D$ denote the merged preference set,

$$D = \{v_i \succ u_i\}_{i=1,2,\ldots,m},$$

where $v_i, u_i$ are taken from $X$. After learning from $D$, the Gaussian process will be able to assign a latent value $f(x)$ to any vector $x \in \mathcal{R}^d$. In addition, the variance of a latent value can be inferred. Bayesian optimization will be performed based on these latent values.

The prior probability of $f(x_i)$ is defined as

$$P(\boldsymbol{f}) = \frac{1}{(2\pi)^{\frac{n}{2}}|\Sigma|^{\frac{1}{2}}} \exp\left(-\frac{1}{2}\boldsymbol{f}^T \Sigma^{-1} \boldsymbol{f}\right),$$

where $\boldsymbol{f} = [f(x_1), f(x_2), \ldots, f(x_n)]^T$, and $\Sigma$ is the covariance matrix defined by a radial basis function kernel [7]. Using Gaussian noise variables $\delta \sim \mathcal{N}(\delta; 0, \sigma^2)$, the probability of preference $v_k \succ u_k$ is described as

$$P(v_k \succ u_k | f(v_k), f(u_k))$$
$$= \int\int P(v_k \succ u_k | f(v_k) + \delta_v > f(u_k) + \delta_u) \mathcal{N}(\delta_v; 0,1) \mathcal{N}(\delta_u; 0,1) d\delta_v d\delta_u.$$

The probability of data generation is then defined as

$$P(D|\boldsymbol{f}) = \prod_{k=1}^{m} P(v_k \succ u_k | f(v_k), f(u_k)).$$

By using Bayes' theorem, we can arrive at the posterior probability,

$$P(\boldsymbol{f}|D) = \frac{P(\boldsymbol{f})P(D|\boldsymbol{f})}{P(D)} = \frac{P(\boldsymbol{f})}{P(D)} \prod_{k=1}^{m} P(v_k \succ u_k | f(v_k), f(u_k)).$$

The maximum a posteriori estimate (MAP) of the latent values is defined as $\boldsymbol{f}_{\text{MAP}} = \arg\max_{\boldsymbol{f}} P(\boldsymbol{f}|D)$. Taking the logarithm of the posterior probability, the solution is obtained by minimizing

$$S(\boldsymbol{f}) = -\sum_{k=1}^{m} \ln \Psi(s_k) + \frac{1}{2}\boldsymbol{f}^T \Sigma^{-1} \boldsymbol{f}$$

where $s_k = \frac{f(v_k) - f(u_k)}{\sqrt{2}\sigma}$ and $\Psi(s) = \int_{-\infty}^{s} \mathcal{N}(\gamma; 0, 1) d\gamma$.

To make a prediction at a new sample point $x^*$, we infer the probability distribution of its latent value as

$$P(f^*|D) = \int P(f^*|\boldsymbol{f}) P(\boldsymbol{f}|D) d\boldsymbol{f} \sim N(f^*; \boldsymbol{K}^{*T} \Sigma^{-1} \boldsymbol{f}_{\text{MAP}}, K^{**} - \boldsymbol{K}^{*T}(\Sigma + \Lambda_{\text{MAP}}^{-1})^{-1} \boldsymbol{K}^*)$$

,

where $\boldsymbol{K}^* = [K(x^*, x_1), K(x^*, x_2), \ldots, K(x^*, x_n)]^T$, $K^{**} = K(x^*, x^*)$ and $\Lambda_{\text{MAP}}$ is the Hessian matrix $\frac{\partial^2 S(\boldsymbol{f})}{\partial \boldsymbol{f} \partial \boldsymbol{f}^T} - \Sigma^{-1}$ at $\boldsymbol{f} = \boldsymbol{f}_{\text{MAP}}$. The predicted mean and variance of the latent value at $x^*$ are $\mu^* = \boldsymbol{K}^{*T} \Sigma^{-1} \boldsymbol{f}_{\text{MAP}}$ and $\sigma^{*2} = K^{**} - \boldsymbol{K}^{*T}(\Sigma + \Lambda_{\text{MAP}}^{-1})^{-1} \boldsymbol{K}^*)$, respectively. All hyperparameters are set as instructed in [9].

2.2 Bayesian optimization based on preference learning

In Bayesian optimization, the mean latent value $\mu^*$ and standard deviation $\sigma^*$ are computed for all remaining candidates. Let $\mu_{max}$ denote the maximum value observed so far. The expected improvement of a candidate $x^*$ is described as follows.

$$\text{EI}(x^*) = (\mu_{max} - \mu^*)\Phi\left(\frac{\mu_{max} - \mu^*}{\sigma^*}\right) + \sigma^*\varphi\left(\frac{\mu_{max} - \mu^*}{\sigma^*}\right),$$

where $\Phi$ and $\varphi$ represent the cumulative distribution function and the probability density function of standard normal distribution, respectively. The candidate with maximum expected improvement is chosen for next observation.

## 3. Results

### 3.1 Absorption wavelength of molecules

Most large-scale public databases provide materials properties obtained from first principle calculations, not experimental ones [1,2]. It is thus interesting to see if computational data can help the search for best materials. We created our own small database of 90 organic molecules with their absorption spectra computed via TD-DFT. The set of molecules in denoted as $A$. See [4] for computational details. In our first benchmark, we examine how much this database can accelerate experimental search of molecules with longest absorption wavelength.

The experimental dataset $C$ contains $N=14$ molecules from our previous publication [4]. We synthesized these molecules and measured absorption wavelength with UV spectroscopy. They are all included in our database, $C \subset A$, but there is a considerably large gap between experimental and computational absorption wavelengths (Supplementary Table 1). We created five types of 'legacy' datasets, each consists of 50 molecules. For $q=0, 25, 50, 75$ and $100$, the *q%-overlap dataset* consists of $\lfloor qN/100 \rfloor$ molecules in $C$, $50 - \lfloor qN/100 \rfloor$ molecules in *A-C*, and their computational wavelengths.

To see how the Gaussian process model is enhanced due to a legacy dataset, we evaluated it with ranking accuracy. First, molecules in $C$ are divided into 80% training set and 20% test set. A Gaussian process model is trained with preferences derived from the training set and a legacy dataset. As descriptors, 200 dimensional features were obtained using RDKit Descriptors Calculators [10,11]. The trained model is used to compute latent values of test examples. For the test set, the difference between two rankings due to experimental wavelengths and latent values are measured with an accuracy measure called NDCG [12]. If rankings are completely identical, NDCG is one. A smaller value of NDCG indicates a larger difference in rankings. Figure 3(a) shows the ranking accuracy without any legacy dataset (i.e., single dataset) and that with a various type of legacy dataset. Each violin plot is created with 50 different training/test splits. The accuracy improved, as the degree of overlap is increased and the accuracy is almost perfect for 100% overlap. The result matched our intuition that a legacy data is more valuable when overlap is larger (Figure 1). The accuracy is enhanced at 0% overlap as well, indicating that a legacy data without overlap can sometimes be of help.

Next, we performed a materials design benchmark using Bayesian optimization. First, two molecules are randomly chosen and the selection with Bayesian optimization is applied from the third molecule. For a degree of overlap, we performed 50 runs of Bayesian optimization, where the initial two molecules and the legacy dataset was resampled in every run. The success rate at iteration $j$ is defined as the fraction of runs where the best molecule was found within $j$ selections of molecules. Figure 3(b) shows the result without legacy set (i.e., single dataset) and with a different type of legacy set. Since our experimental dataset $C$ was very small, the performance for single dataset was poor. Improvement with a legacy data was observed at all cases in including 0% overlap, indicating that preference learning can retrieve useful information from legacy data without explicit calibration.

3.2 Bandgap of inorganic materials

The same series of benchmarking experiments is applied to another subject. The online material database of the National Renewable Energy Laboratory (NREL, https://materials.nrel.gov) provides bandgap calculated by Perdue Burke Ernzerhof (PBE) method of 2142 oxides [13]. Among these oxides, 194 oxides have the bandgap data by many-body GW calculation method [3]. GW calculation predicts band gaps more accurately but is far more computationally expensive than PBE [14,15]. We define a search problem of finding the oxide with largest bandgap in terms of GW. The candidate set $C$ is determined as the 194 oxides with GW bandgaps and the total set $A$ corresponds to the 2142 oxides. Legacy datasets of size 200 are created at different degrees of overlap. 132 dimensional descriptors are obtained using Elementproperty Featurizer of Matminer [16].

Ranking accuracy and Bayesian optimization performances are shown in Figures 4(a) and (b), respectively. With a legacy dataset without overlap, ranking accuracy was worse than that of single dataset. Nevertheless, Bayesian optimization was accelerated in comparison to the single dataset case. As Section 3.1, a larger overlap resulted in higher accuracy and better acceleration.

# 4. Discussion and conclusion

We reported that preference-learning-based data integration works excellently in two kinds of materials datasets. This result is surprising and encouraging at the same time, because the conversion of numerical data to preferences incurs information loss in trade with calibration-free integration. Our method extends easily to deal with more than three datasets. In current materials science, data sharing is not commonly done due to difficulty of integration. Our method may promote cooperation among researchers to save the cost of expensive and time-consuming experiments.

In materials sciences, there is wide-spread misunderstanding that machine learning always require a large amount of data. One favorable aspect of our results is that our method worked in small data scenarios (i.e., less than several hundred data points). When users want to use larger datasets, current implementation of our algorithm may not be very scalable, because the computational complexity is $O(m^3)$ [9] where $m$ is the number of preference relations. Recent developments in Gaussian process and preference learning [17,18] may be beneficial in improving scalability.

# Acknowledgements

This work was supported by ``Materials Research by Information Integration" Initiative (MI2I) project. X.S. would like to gratefully acknowledge the financial support from the China Scholarship Council (CSC NO. 201809120018). K.T. is supported by NEDO P15009, SIP (Technologies for Smart Bio-industry and Agriculture), JST CREST JPMJCR1502, and JST PRISM JPMJCR18Y3. R.T. is supported by JST CREST JPMJCR17J2 and SIP ("Materials Intergration" for Revolutionary Design System of Structural Materials). The authors thank Diptesh Das, Koki Kitai, Jinzhe Zhang and Yuan Yao for discussions.

**Figure 1.** In materials design with a legacy dataset, we search the best one from a set of candidates (red), using the information from a set of examples in the legacy dataset (blue). If these two sets have large overlap (right), we can make most of the legacy data for accelerating the search, while it would be difficult without no overlap (left).

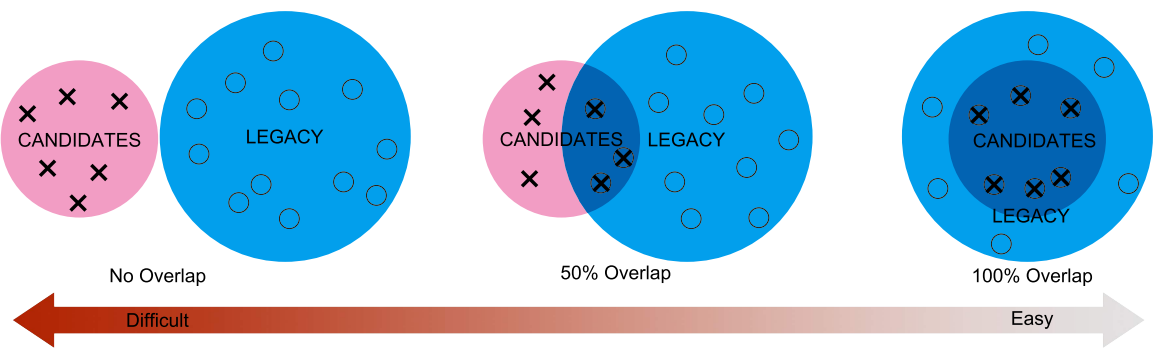

**Figure 2.** Data integration with preference learning. Two experimental datasets measuring the same property are available but the measured values have a gap due to different conditions. In our method, each dataset is separately translated to preference relations. A Gaussian process model is trained from all the relations. The trained model yields probability distributions of latent values at all points in the descriptor space.

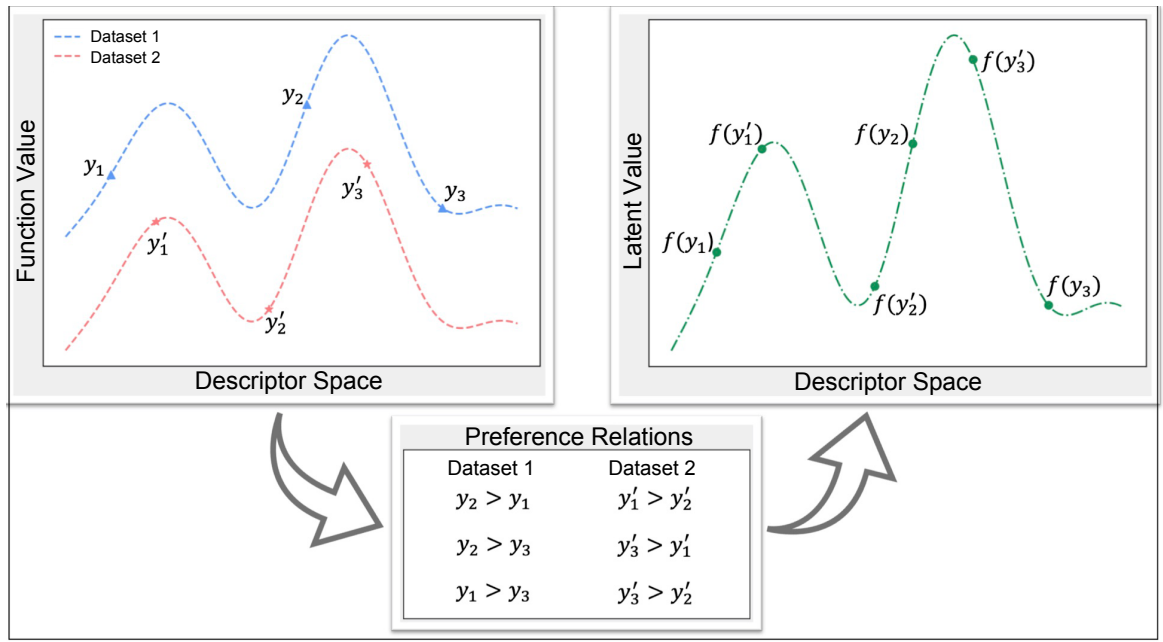

**Figure 3.** Results for organic molecules. (a) Ranking accuracy by Gaussian process with preference learning. (b) Success rate of Bayesian optimization with preference learning against the number of iterations.

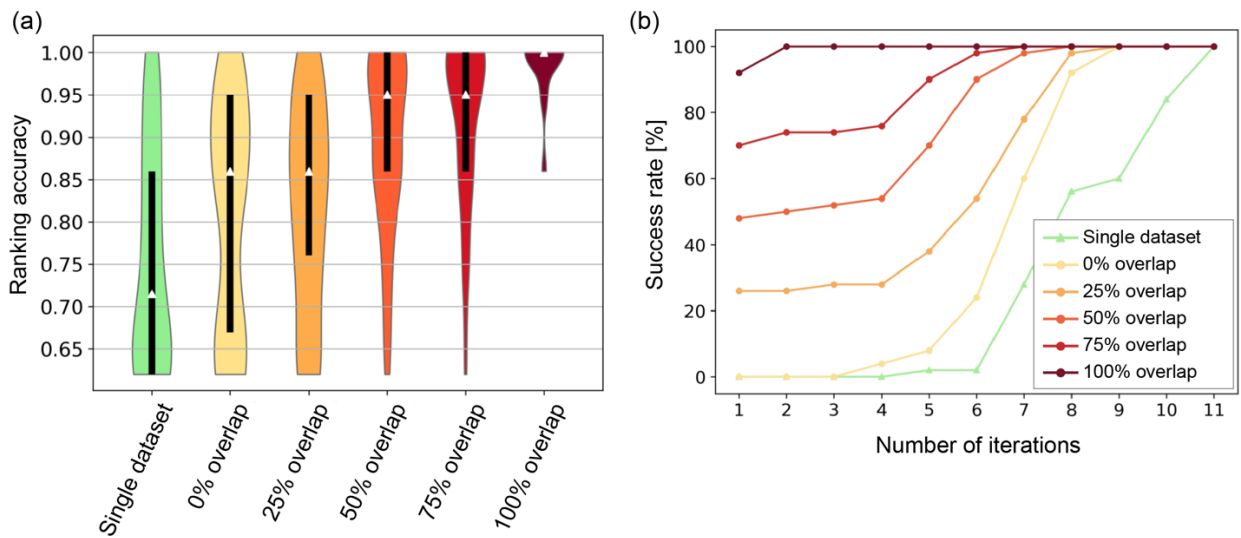

**Figure 4.** Results for oxides. (a) Ranking accuracy by Gaussian process with preference learning. (b) Success rate of Bayesian optimization with preference learning against the number of iterations.

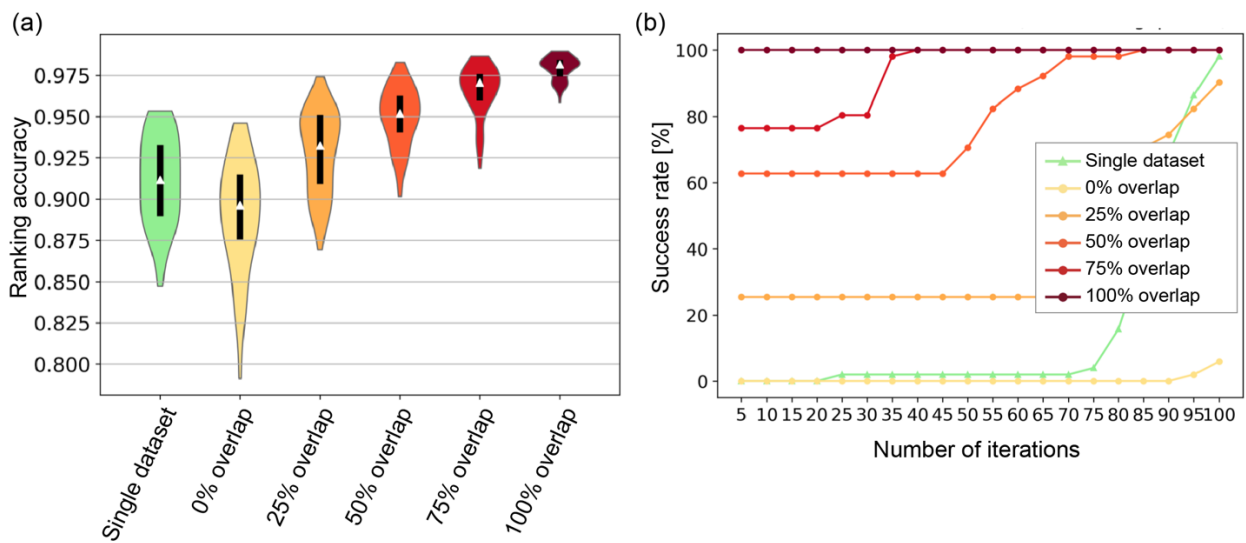

Supplementary material: Leveraging Legacy Data to Accelerate Materials Design via Preference Learning

Xiaolin Sun, Zhufeng Hou, Ryo Tamura, Masato Sumita, Shinsuke Ishihara, Koji Tsuda

Supplementary Table 1: SMILES strings of molecules and their experimental and computational absorption wavelengths.

| Molecule(SMILES) | Experimental/nm | Computational/nm |
|---|---|---|
| Cc1occn1 | 196.0 | 207.83 |
| Cc1ncc(n1C)O | 334.0 | 207.42 |
| Cc1ccnc2c1cc(O)cc2 | 332.0 | 301.57 |
| Oc1ccc2c(c1)cccc2C | 331.0 | 295.92 |
| O=NN(Cc1ccccc1O)C | 411.0 | 400.81 |
| CC(=O)C(=O)CN(C)C | 470.0 | 512.69 |
| COc1ccc(NC2=C(Cl)C(=O)c3ccccc3C2=O)cc1 | 489.0 | 581.15 |
| CC(=O)Nc1ccc(NC2=C(Cl)C(=O)c3ccccc3C2=O)cc1 | 491.0 | 569.36 |
| O=C1OC(/C=C/c2ccccc2)=N/C1=C/c1ccc(Cl)cc1 | 384.0 | 408.66 |
| COc1nc(Nc2ccc(-c3nc4ccccc4o3)cc2)nc(OC)n1 | 324.0 | 322.15 |
| C=C1N(C2CCCCC2)C(=O)OC12CCCCC2 | 222.0 | 204.83 |
| CC1OC(C)OC(C)O1 | 180.0 | 133.5 |
| CN(C)c1ccc(NC2=C(Cl)C(=O)c3ccccc3C2=O)cc1 | 552.0 | 692.99 |
| O=[N+]([O-])c1ccc(/C=N/c2ccc(N3CCOCC3)cc2)s1 | 460.0 | 500.26 |
| NC(CCC#N)O | - | 187.9 |
| OCNN/C=N/O | - | 214.61 |
| OC1=NCC2(C1)CCCC2 | - | 210.64 |
| CNC[C@@H](C(=O)O)O | - | 216.14 |
| N[C@@H](C[C@H](CC(C)C)O)Cc1ccco1 | - | 218.76 |
| Cc1onc(c1)O | - | 200.19 |
| N[C@H](/C(=NO)/O)CCC | - | 217.61 |
| ON1CC1 | - | 191.69 |
| O[C@H]([C@@H]1CCNCC1)N(C) | - | 189.79 |
| NC[C@H]1OC[C@H]([C@H]([C@H]1O)C)O | - | 212.47 |
| N[C@@H]([C@@H](CC(O)C)O)Cc1cnc[nH]1 | - | 203.28 |
| C1OCN1CN1CCOCC1 | - | 202.96 |
| O[C@@H]([C@H]([C@H](CN)C)O)ON(CC)CC | - | 219.52 |
| N[C@H](CCN1CCNCC1)C | - | 197.73 |
| N[C@H](CC#CC(C)C)O | - | 185.7 |
| OCCCCN(CCO)C[C@@H](O)C | - | 184.41 |
| ON=C(O)C | - | 205.47 |
| C/C=N/N1CC[C@H](C1)O | - | 211.44 |
| C/C=N/N[C@H]1CCCCO1 | - | 207.96 |
| NC(C)(C)C | - | 180.7 |
| ONCCC[C@H](CC(C)C)O | - | 185.83 |
| C1OCN1 | - | 204.52 |
| O[C@@H]1CN2CC[C@H]1CC2 | - | 185.49 |
| C1NCCOCC1 | - | 182.76 |
| CCON/C(=NC)/O | - | 218.04 |
| OC[C@@H](NC[C@H](O)C)O | - | 187.37 |
| OC[C@@H](OCCCN(C)C)C | - | 183.27 |
| OC[C@@H]([C@H]([C@@H]([C@@H](O)C(=N)O)O)O)O | - | 188.61 |
| NN1C(=N)OC[C@H]1C | - | 181.22 |
| O[C@H]1C[C@H]2C([C@@H](C1)N2C)O | - | 195.73 |
| C=C[C@@H]1CCC(=N1)O | - | 213.01 |
| C1OC[C@@H]2N(C1)CCO2 | - | 187.11 |
| N[C@@H]1C(=O)[C@@]2(C([C@H]1CC2)(C)C)C | - | 299.81 |
| N#Cc1c(OC)cc[nH]c1=O | - | 300.7 |
| C/N=C(/O[N][CH]c1ccc(cc1)OC)O | - | 282.13 |
| NN/C(=Cc1ccccc1)/O | - | 294.84 |
| C/N=C(c1n(CC)cnc1/C(=NCC)/O)/O | - | 306.8 |
| Nc1cc(ccc1C)c1ccc(c(c1)O)N | - | 284.09 |
| Oc1nc(c(o1)c1ccccc1)N | - | 287.15 |
| Oc1cn(c(c1)C(=O)O)C(=O) | - | 307.35 |
| COc1nc(C)nc(c1)n1ccccc1=O | - | 315.74 |
| ON1[CH]C(=C1)C(=O)[O] | - | 280.07 |
| C/N=C(c1ccccn1)/O | - | 296.27 |
| O/N=C/1C=Cc2c(C1)cccc2 | - | 307.78 |
| NN(=O)=O | - | 302.1 |
| NN/C(=Nc1ccccc1)/OC(=O)C | - | 300.84 |
| Cc1cc(no1)CCC=O | - | 306.55 |
| O=Cc1c(nn(c1O)C)C | - | 280.32 |
| C/C(=Nc1ccccc1/N=C(/O)C)/O | - | 286.12 |
| C/C=C/C(=NCCN1CCN(C1=O)O)/O | - | 290.94 |
| OC[C@@](C(=O)C)(N)C | - | 286.24 |
| CCc1cccc2c1nccc2 | - | 302.35 |
| O=c1[nH]cccn1 | - | 315.92 |
| NN[C@@H](C(=O)O)CC(=O)O | - | 299.2 |
| ON1C(=O)CC2(C1=O)CCN(CC2)C | - | 318.59 |
| ONc1nc(=O)c2c([nH]1)cccc2 | - | 304.62 |
| Cc1c[nH]c(c1)c1ccc(o1)N(=O)=O | - | 392.77 |
| CC1CC(=O)N(C(=O)C1=O)C | - | 398.56 |
| O=C(c1ccc(cc1)C)/C=C/c1ccccn1 | - | 394.46 |
| O[C@H](Cn1ccnc1N(=O)=O)OC(C)C | - | 388.38 |
| N#Cc1c(C)ccnc1N(=O)=O | - | 417.55 |
| N[C@@H]1ON=C(C1=O)O | - | 418.02 |
| O[C@@H](C([C@H](c1ccccc1)C)N)N(N=O)C | - | 398.9 |
| COc1c(ccc(c1)C)N(=O)=O | - | 389.88 |
| O=NN1CC/C(=C1)/[C@]1(CCCCC1)CN1CCCCC1 | - | 416.93 |
| OC(=O)/C(=C/c1ccccc1)/C(=O)C | - | 400.63 |
| N[N]C1=C[CH]C(=CN1)OC | - | 401.46 |
| N[N]C1=C[CH]C(=C)CN=C1O | - | 380.21 |
| [O]N1[CH]Cc2c(C1)cccc2 | - | 480.46 |
| [O][N]N1[CH]N=C([N]1)NN(=O)=O | - | 483.83 |
| [O][N]N(c1ccccc1)C(=O)c1ccccc1 | - | 487.75 |
| [O][N]O/C(=NCC)/N | - | 484.53 |
| [O][N]O/C=N/c1ccccc1 | - | 500.24 |
| [O][N]O/C(=NCC)/O | - | 500.23 |
| [O][N]N1[CH]N=C([N]1)NN(=O)=O | - | 489.31 |
| [O][N]N1[CH]N=C([N]1)N | - | 486.36 |
| [O][N]N1[CH]N=C([N]1)O | - | 487.37 |
| [O]N(N(c1ccccc1)[O])c1cccc(c1)N | - | 484.17 |
| [O][N]N1[C@@H](CCN=C1O)Cc1ccccc1 | - | 482.01 |
| O=Nn1c(O)nccc1=O | - | 606.58 |